\documentclass[12pt,showpacs,showkeys,amssymb,amsmath]{revtex4}
\usepackage{epsfig}
\usepackage{graphicx}
\usepackage{epic}
\usepackage{eepic}
\usepackage{bbold}

\newcommand{\be}{\begin{equation}}
\newcommand{\ee}{\end{equation}}
\newcommand{\ba}{\begin{eqnarray}}
\newcommand{\ea}{\end{eqnarray}}
\newcommand{\RefEq}[1]{{(\ref{#1})}}

\begin{document}

\title{ Phase--Space Structure for Narrow Planetary Rings }

\author{Luis Benet} 
\email{benet@fis.unam.mx}
\address{Centro de Ciencias F\'{\i}sicas, U.N.A.M., Apdo. Postal 48--3, 62251
  Cuernavaca, Mor., M\'exico}

\author{Olivier Merlo} 
\email{merlo@quasar.physik.unibas.ch}
\address{Institut f\"ur Physik, Universit\"at Basel, Klingelbergstr. 82,
  CH--4056 Basel, Switzerland}

\date{\today}

\begin{abstract}
  We address the occurrence of narrow planetary rings under the interaction
  with shepherds. Our approach is based on a Hamiltonian framework of
  non--interacting particles where open motion (escape) takes place, and
  includes the quasi--periodic perturbations of the shepherd's Kepler motion
  with small and zero eccentricity. We concentrate in the phase--space
  structure and establish connections with properties like the eccentricity,
  sharp edges and narrowness of the ring. Within our scattering approach, the
  organizing centers necessary for the occurrence of the rings are stable
  periodic orbits, or more generally, stable tori. In the case of eccentric
  motion of the shepherd, the rings are narrower and display a gap which
  defines different components of the ring.
\end{abstract}



\maketitle

\section{Introduction}

Narrow rings have impressed us ever since their observation by Galileo. With
the discovery of the rings of Uranus in 1977 by stellar--occultation
measurements, the view of the rings as a special feature of Saturn changed
completely. This went on with the information obtained in particular by the
Voyager missions, which provided for the first time a close view into the
complexity that takes place in the ring systems. We now know that rings appear
in all giant planets. They display different morphologies, sizes, masses and
physical properties. Different relevant physical phenomena takes place on
them: Surface--wave phenomena, electromagnetic forces acting on micrometric
charged dust grains and gravitational confinement due to nearby satellites are
some examples. Different aspects of ring structure, dynamics and open problems
are reviewed in Refs.~\cite{rings,Nicholson91,Esposito02}. New information is
becoming available nowadays by the Cassini mission.

After the discovery of Uranus' rings, which are extremely narrow in comparison
to those of Saturn, models including confinement due to nearby moons were
introduced. In particular, we mention the shepherding model introduced by
Goldreich and Tremaine~\cite{Goldreich79}, where two (shepherd) moons around
the Uranian rings are proposed to bound and to define the sharp edges of the
rings. The mechanism to maintain the ring involves angular momentum transfer
between the shepherd moons and the ring particles as well as viscous damping
due to interparticle collisions. While the full scenario for shepherding has
not been fully understood, the presence of dissipation seems to be essential
for the argument, which is in addition formulated assuming that the ring
boundary is located at a lower--order resonance (see~\cite{Esposito02}). In
this paper, we propose a purely Hamiltonian formulation of the shepherding
mechanism that confines the ring, focusing in particular in the phase--space
structure that characterizes it. We do this by assuming, in a first
approximation, an independent--particle model, i.e., where the ring particles
do not interact among their selves. Within our approach, we succeed in
explaining the confinement, on the one hand, and also long standing problems
like the narrowness of the ring and its sharp edges.

We choose a rather simple but unrealistic example to illustrate the relevant
structures in phase space and understand its implications. We emphasize that
the example we define here, which is based on a billiard system with an
arbitrary large number of non--interacting particles, is simply a rough
cartoon of the real interaction between the particles of the ring and the
shepherd moons. We are not proposing by any means that the planetary rings are
due to real collisions between the particles of the ring and the shepherds, or
that the interactions among particles of the ring are unimportant. Our goal is
to point out that, within the independent particle picture considered here,
the conclusions we can draw from the phase--space structure are {\it generic}
and therefore of interest in the context of planetary rings. The reason is
that the relevant phase--space structures are independent of the precise
details of the system. This is the reason we cautiously refer to it as an {\it
  example} or {\it toy} model rather than a {\it model} for the (shepherded)
narrow rings. The reader should not be misguided by this, which simply avoids
the complications of working explicitely with $1/r$ potentials.

The paper is organized as follows: In Sec.~\ref{sec_toy} we define the toy
model within a Hamiltonian formulation, using an inertial reference frame and,
what turns to be more general and convenient, using rotating--pulsating
coordinates. We consider only the influence of what it would be interpreted as
one shepherd moon for simplicity. In Sec.~\ref{sec3} we consider the
occurrence of rings in the case of circular motion of the shepherd. New
results including analytical estimates of the bounds of the ring are presented
here. In Sec.~\ref{sec4} we consider the case of small but non--zero
eccentricity, and show how the arguments carry on in this case. Here, we
present indirect evidence for the existence of trapped motion in the system
under consideration. Finally, in Sec.~\ref{concl} we present our conclusions
and some outlook of the work.

\section{The {\it toy} model}
\label{sec_toy}

\begin{figure}
  \includegraphics[angle=90,width=8.5cm]{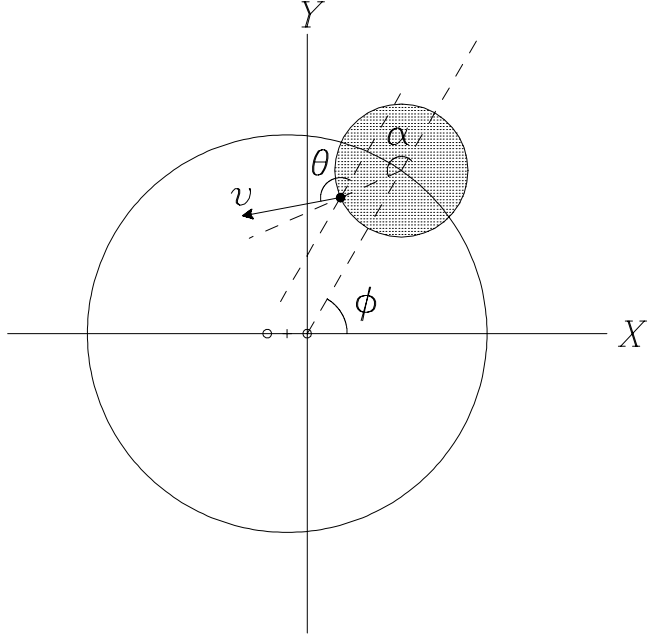}
\caption{\label{fig1}
  Geometry of the toy model: $\phi$ denotes the initial position of the center
  of the disk; $\alpha$ denotes the position of collision point on the disk;
  $v$ denotes the magnitude of the outgoing velocity and $\theta$ defines its
  direction. The foci of the ellipse described by the center of the disk are
  shown on the $X$--axis as open circles ($\circ$), and the center by a cross
  ($+$).}
\end{figure}

Consider the planar motion of a massless point particle off {\it one} circular
hard--disk of radius $d$ which moves on a two--dimensional keplerian periodic
orbit (Fig.~\ref{fig1}). The potential is therefore zero everywhere except at
the position occupied by the disk, where it is infinite. The center of the
disk describes an elliptic Kepler orbit with one focus at the origin. This
example is a rough cartoon of the restricted three--body
problem~\cite{Meyer95}: In the present case it is defined by the central
planet, one shepherd moon orbiting the planet and one particle of the ring. We
restrict ourselves here to the three--body version of the problem to simplify
the discussion. Results for the four--body case will be briefly discussed in
Sec.~\ref{concl}.

The radial position of the center is denoted by $R(\phi)$, where $\phi$
defines the position along the Kepler orbit measured from the pericenter.
$R(\phi)$ is given by the usual expression in polar coordinates in terms of
the semi--mayor axis $a=1$ and the eccentricity $\varepsilon$. We shall focus
on the case of small or zero eccentricity, since this is the situation
commonly encountered. For non--zero $\varepsilon$, due to the explicit time
dependence the system has two--and--half degrees of freedom and no constant of
motion. Hence, the phase space is five dimensional: Two coordinates and their
canonically conjugated momenta define the phase--space coordinates of the
particle, and the angle $\phi$ determines the position of the disk along the
Kepler orbit. In the circular case, as we show below, there is a constant of
motion which reduces the dimensionality of phase space.

The Hamiltonian, expressed in an inertial reference frame, can be written as
\be
\label{eq1}
H = \frac{P_X^2+P_Y^2}{2}+V(|\vec{X}-\vec{X}_d(\phi)|).
\ee
Here, $\vec{X}_d$ denotes the position vector of the center of the disk, which
implies an explicit time--dependent potential through the dependence upon
$\phi$, and $\vec{X}$ is the position of the particle. The potential
$V(|\vec{X}-\vec{X}_d(\phi)|)$ is zero for $|\vec{X}-\vec{X}_d(\phi)|^2>d^2$
and infinite otherwise. We perform a canonical transformation to
rotating--pulsating coordinates by means of the generating function
\be W = \frac{R(\phi)}{\bar{R}} \left[P_X (x\cos\phi-y\sin\phi) + P_Y
  (x\sin\phi+y\cos\phi)\right] \nonumber, 
\ee
where $\bar{R}$ is the mean--orbital radius of the disk. In this coordinates,
the disk is at rest at $\vec{x}_d=(\bar{R},0)$. The new Hamiltonian is then
given by
\be
\label{eq2}
J = \frac{1}{2}\frac{\bar{R}^2}{[R(\phi)]^2} (p_x^2+p_y^2)
  - \dot\phi(xp_y-yp_x) 
  - \dot\phi\frac{1}{R(\phi)}\frac{{\rm d}R(\phi)}{{\rm d}\phi}(xp_x+yp_y) 
  + V(|\vec{x}-\vec{x}_d|).
\ee
Note that in the case of a circular orbit we have $R=a$ and $\phi=\omega t$,
where $\omega=\dot\phi$ is the orbital frequency and $t$ is the time. In this
case, the third term of the r.h.s. of the Hamiltonian~\RefEq{eq2} vanishes. By
consequence, $J$ is time independent, and thus a constant of motion. This is
the so--called Jacobi integral of the restricted three--body problem. We shall
refer to the value of $J$ in Eq.~\RefEq{eq2} as the Jacobi integral, even in
the case of non--vanishing eccentricity. In the following we shall drop the
potential in Eq.~\RefEq{eq1} and~\RefEq{eq2}, and restrict implicitly to the
case $|\vec{X}-\vec{X}_d(\phi)|^2 = 
 (R(\phi)/\bar{R})^2 |\vec{x}-\vec{x}_d|^2 >d^2$.

The dynamics of this scattering system are straightforward. The particle moves
freely on a rectilinear trajectory of constant velocity until it encounters
the disk. No collision leads to open motion, i.e., the particle escapes away
from the interaction region. In the context of the ring, this situation is
interpreted as the case of a particle abandoning the bulk of the ring. If a
collision takes place, the particle is specularly reflected with respect to
the local (moving) frame of the disk at the collision point. This defines the
outgoing conditions after the collision, and then the motion is again
rectilinear uniform~\cite{Meyer95}. The precise result of a collision depends
on the position where the collision occurs on the disk, the relative
velocities, and for non--vanishing $\varepsilon$, on the position along the
elliptic trajectory of the disk. Collisions taking place on the front of the
disk increase the (outgoing) kinetic energy of the particle, while collisions
on the back reduce it.

In the context of the rings, we are interested in (ring) particles which are
dynamically trapped; within the present example, this can only be through
consecutive collisions with the disk. A convenient description of this
situation is given by introducing the following quantities, which are defined
at the collision point with the disk (cf. Fig.~\ref{fig1}). The angle $\phi$
characterizes the position of the center of the disk with respect to the $X$
axis (inertial frame), and $v$ is the magnitude of the velocity after the
collision (outgoing velocity). The angle $\alpha$ denotes the angle formed by
$\vec{X}_d(\phi)$ and the position vector of the collision point referred to
the center of the disk. Finally, the angle $\theta$ defines the outgoing
direction of the velocity.

\section{Occurrence of rings: The circular case}
\label{sec3}

The dynamics of the disk define naturally a map at the collision point with
the disk. This map is {\it open} in the sense that certain initial conditions
may not have an associated image. This situation corresponds to the case where
the particle escapes after a given collision. On the other hand, trapped
trajectories are associated with consecutive collisions with the disk,
irrespective whether such trajectories display periodic, quasi--periodic or
chaotic behavior. Periodic orbits and their stability are thus of central
interest in the context of the rings. In particular, simple periodic orbits
for the circular case can be worked out analytically for the example
considered here.

Between collisions, i.e. during the free motion of the particle, the linear
equations of motion in the rotating frame can be solved explicitely. The
Jacobi integral is then given by ($\omega=1$)
\be
\label{eq3}
J = \frac{v^2}{2}-v\left(R\sin\theta + d\sin(\theta-\alpha)\right).
\ee

We notice that the radial collisions ($\alpha=\pi$) do conserve the kinetic
energy, and are thus fixed points of the dynamical map in the rotating frame.
Starting from the initial conditions of a radial collision, in order to have
that the next collision is radial too the velocity must be given by
\be
\label{eq4}
v = - \frac{2 (R-d) \cos\theta}{\Delta\phi} = - \frac{2 (R-d)
  \cos\theta}{(2n-1)\pi+2 \theta}.
\ee
Here, $n=0,1,\dots$ denotes the number of full turns completed by the disk
between consecutive radial collisions, and $\Delta\phi= (2n-1)\pi+2 \theta$ is
the corresponding change in the angle $\phi$. For these orbits, the Jacobi
integral is given by
\be
\label{eq5}
J_n = (R-d)^2 \frac{2\cos^2\theta+\Delta\phi\sin(2\theta)}{(\Delta\phi)^2} =
(R-d)^2 \frac{2\cos^2\theta(1+\Delta\phi\tan\theta)}{(\Delta\phi)^2}.
\ee

\begin{figure*}
  \includegraphics[angle=90,width=16cm]{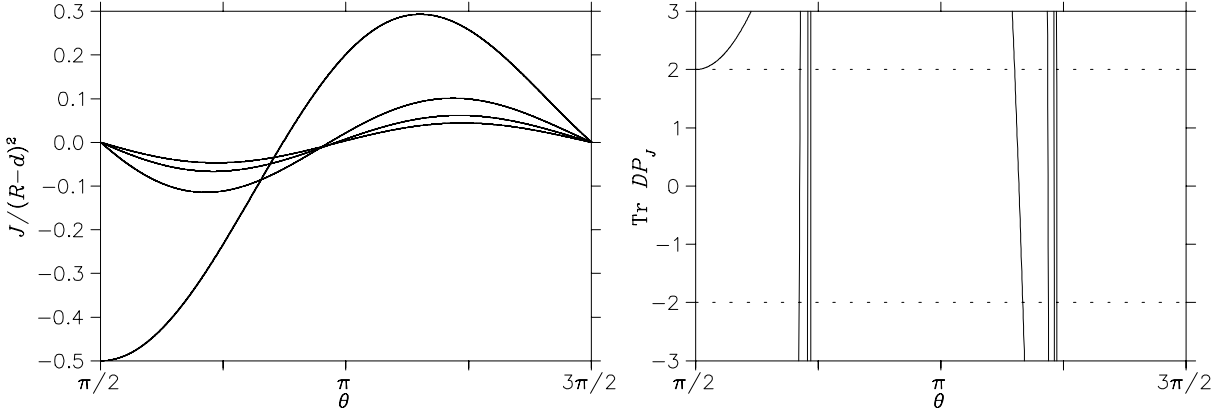}
\caption{\label{fig2} 
  Consecutive radial collision orbits (simple periodic orbits) of the circular
  rotating billiard for $n=0,1,2,3$: (a)~Jacobi integral $J$ in units of
  $(R-d)^2$ and (b)~${\rm Tr\,} D{\cal P}_J$ in terms of $\theta$.}
\end{figure*}

In Fig.~\ref{fig2}a we illustrate the structure of $J_n/(R-d)^2$, for some
values of $n$. Note that the curves display one maximum and one minimum for
each value of $n$. Moreover, for negative values of $J$, two periodic orbits
can have the same values of $\theta$ and the Jacobi integral, while the
characteristic $n$'s are different. This situation is a consequence of the
quadratic character of Eq.~\RefEq{eq3}, which for $J<0$ has two distinct
solutions for $v$. The occurrence of maxima and minima on these curves
suggests the appearance of these consecutive collision orbits by
saddle--center bifurcations.

For $\varepsilon=0$, Eq.~\RefEq{eq2} defines a time--independent two degrees
of freedom system. In this case, the stability of the periodic orbits can be
checked by standard procedures involving the trace and the determinant of the
linearized matrix of the Poincar\'e map at the fixed points. The
transformation $(\alpha_{n+1}, p_{n+1})={\cal P}_J(\alpha_n,p_n)$ with
\be
\label{eq6}
p_n=-d-R\cos\alpha_n-v\sin(\alpha_n-\theta_n),
\ee
defines a canonical Poincar\'e map. Here, $\alpha_n\in [0,2\pi]$ and
$p_n\in[-p_{max},p_{max}]$, where $p_{max}=(2J+R^2+d^2+2Rd\cos\alpha)^{1/2}$.
The determinant of the linear approximation is thus $1$. [The same map but
with $p_n=\theta$ which was used in Ref.~\cite{Meyer95} has also unit
determinant for the consecutive collision periodic orbits, even--though it is
non--canonical.] The information on the stability is thus completely contained
in the trace of the linearized matrix $D{\cal P}_J$. One finds~\cite{Merlo04}
\be
\label{eq7}
{\rm Tr\,} D{\cal P}_J = 2 +
\frac{(\Delta\phi)^2(1-\tan^2\theta)-4(1+\Delta\phi\tan\theta)}{d/R}.
\ee
It is well known that changes in the stability of the periodic orbits
(bifurcations) take place when the characteristic multipliers are $\pm 1$;
this is equivalent to have ${\rm Tr\,} D{\cal P}_J =\pm 2$. Equating the
r.h.s. of Eq.~\RefEq{eq7} to $2$ turns to be equivalent to the condition ${\rm
  d}J_n/{\rm d}\theta=0$~\cite{Benet00}, i.e., the condition defining the
position of the minima and maxima of $J_n$. The corresponding condition
obtained for $-2$ yields a condition related to period--doubling bifurcations;
we shall come back to this later. Note that ${\rm Tr\,} D{\cal P}_J = 2$ is
independent of $d/R$, while in the case ${\rm Tr\,} D{\cal P}_J = -2$ there is
an explicit dependence on $d/R$. Figure~\ref{fig2}b shows the behavior of
${\rm Tr\,} D{\cal P}_J$ in terms of $\theta$.

\begin{figure*}
  \includegraphics[angle=270,width=7cm]{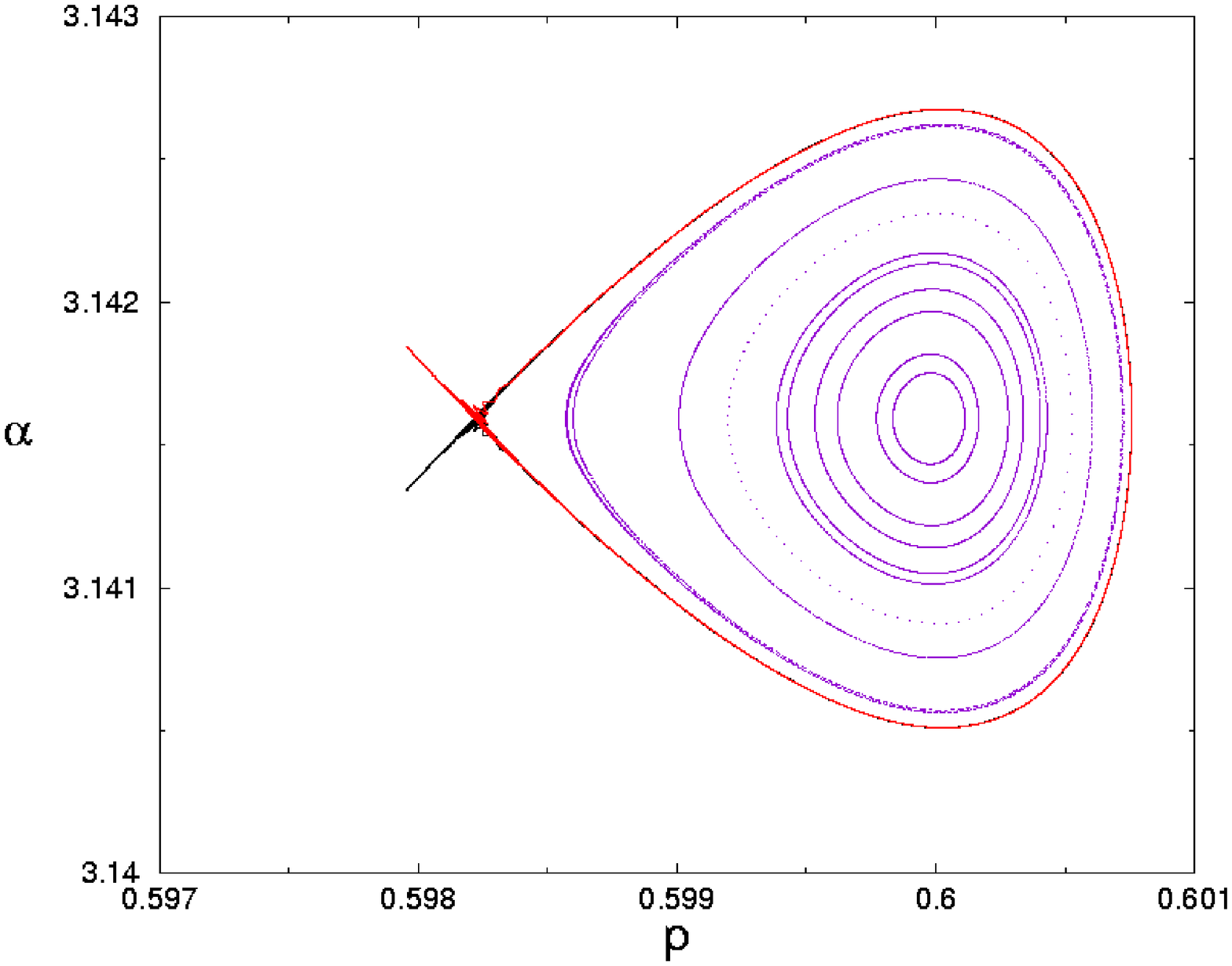} \hspace{0.5cm}
  \includegraphics[angle=270,width=7cm]{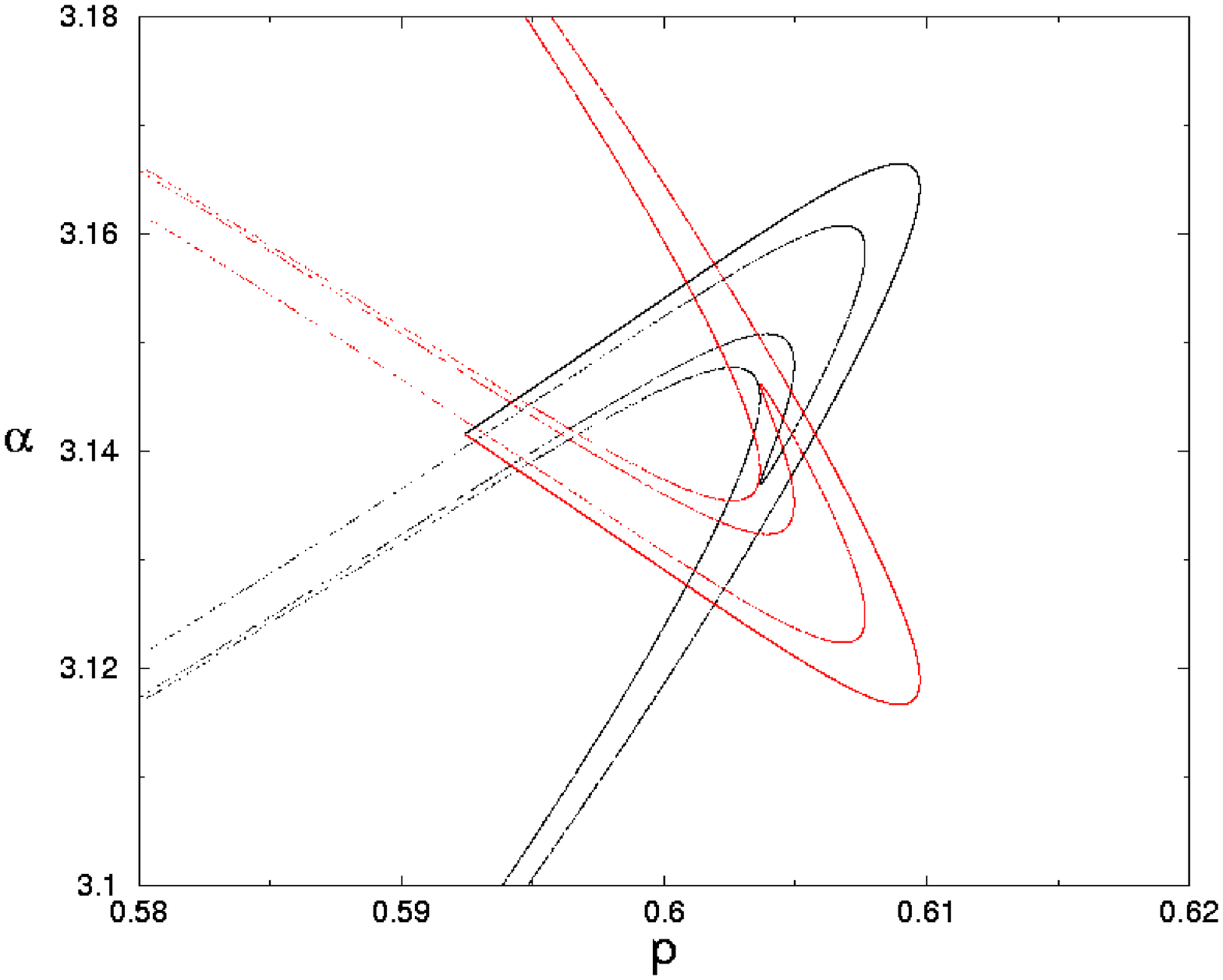}
\caption{\label{fig3} 
  Phase--space structure of the scattering billiard on a circular orbit ($R=1$
  and $d=1/2$). (a)~Situation with one elliptic fixed point
  ($J/(R-d)^2=0.29325$). (b)~Case in which the elliptic fixed point has turned
  into inverse hyperbolic ($J/(R-d)^2=0.29218$). The phase space is dominated
  by unstable motion, although the Smale horseshoe is not complete.}
\end{figure*}

We observe from Fig.~\ref{fig2}b that there are connected intervals in
$\theta$, and therefore in $J$, where radial collision periodic orbits are
strictly stable, i.e. $2>{\rm Tr\,} D{\cal P}_J >-2$. For $J$ within these
intervals of Jacobi integral, the phase--space structure of the scattering
system displays one elliptic fixed point, surrounded by the typical KAM tori
and some chaotic layers (Fig.~\ref{fig3}a). This region in phase space is
bounded by the stable and unstable invariant manifolds of the companion
hyperbolic fixed point, which reach the incoming and outgoing asymptotes of
the scattering system. The motion of particles whose initial conditions lie
within this region display consecutive collisions and therefore are
dynamically trapped. We shall refer to this region as the region of trapped
motion. Note that if the initial conditions of the particle do not belong to
any region of trapped motion, the particle will eventually escape from the
interaction region.

We consider the case $J>0$ for concreteness. By further reducing the value of
the Jacobi integral, there is a value where ${\rm Tr\,} D{\cal P}_J =-2$ is
fulfilled. There, the elliptic fixed point becomes inverse hyperbolic, and a
period doubling bifurcation sets in: The region of trapped motion vanishes
rapidly after further reducing $J$. In Fig.~\ref{fig3}b we have illustrated
this case plotting the phase--space structure corresponding to an incomplete
Smale horseshoe~\cite{Rue94}. We remark that the dependence upon $d/R$ of
${\rm Tr\,} D{\cal P}_J =-2$ implies that the actual parameters of the
billiard define the thickness of the ring. Physically this implies that the
parameters related with the motion and mass of the shepherd moons do indeed
influence the width of the ring. Further reducing the value of $J$ yields a
hyperbolic Smale horseshoe. A detailed description of the complete bifurcation
scenario for this billiard, as well as the dynamics for $J=0$, can be found in
Ref.~\cite{Merlo04}.

We turn now to the rings~\cite{Benet00}. Within the present toy model,
consider an arbitrary large number of {\it non--interacting} particles. The
initial conditions of these particles are completely arbitrary, i.e. have no
restrictions at all (e.g., in the value of the Jacobi integral). Then, the
particles may collide with the disk at any value $\phi$ along the circular
orbit of the disk. We are thus considering the most general ensemble of
initial conditions. Letting the system evolve most particles escape after few
collisions. Yet, those whose initial conditions belong to the phase--space
regions which are trapped by consecutive collisions with the disk will not
escape. This situation may correspond to the initial conditions of the
unstable periodic orbits, or to homoclinic or heteroclinic orbits. These
initial conditions are of no interest for us since they are a set of measure
zero. On the other hand, initial conditions corresponding to values of $J$
with a phase--space structure like in Fig.~\ref{fig3}a, have in addition the
important property of having a small but finite, i.e. non--zero, measure.

\begin{figure*}
  \includegraphics[angle=270,width=7cm]{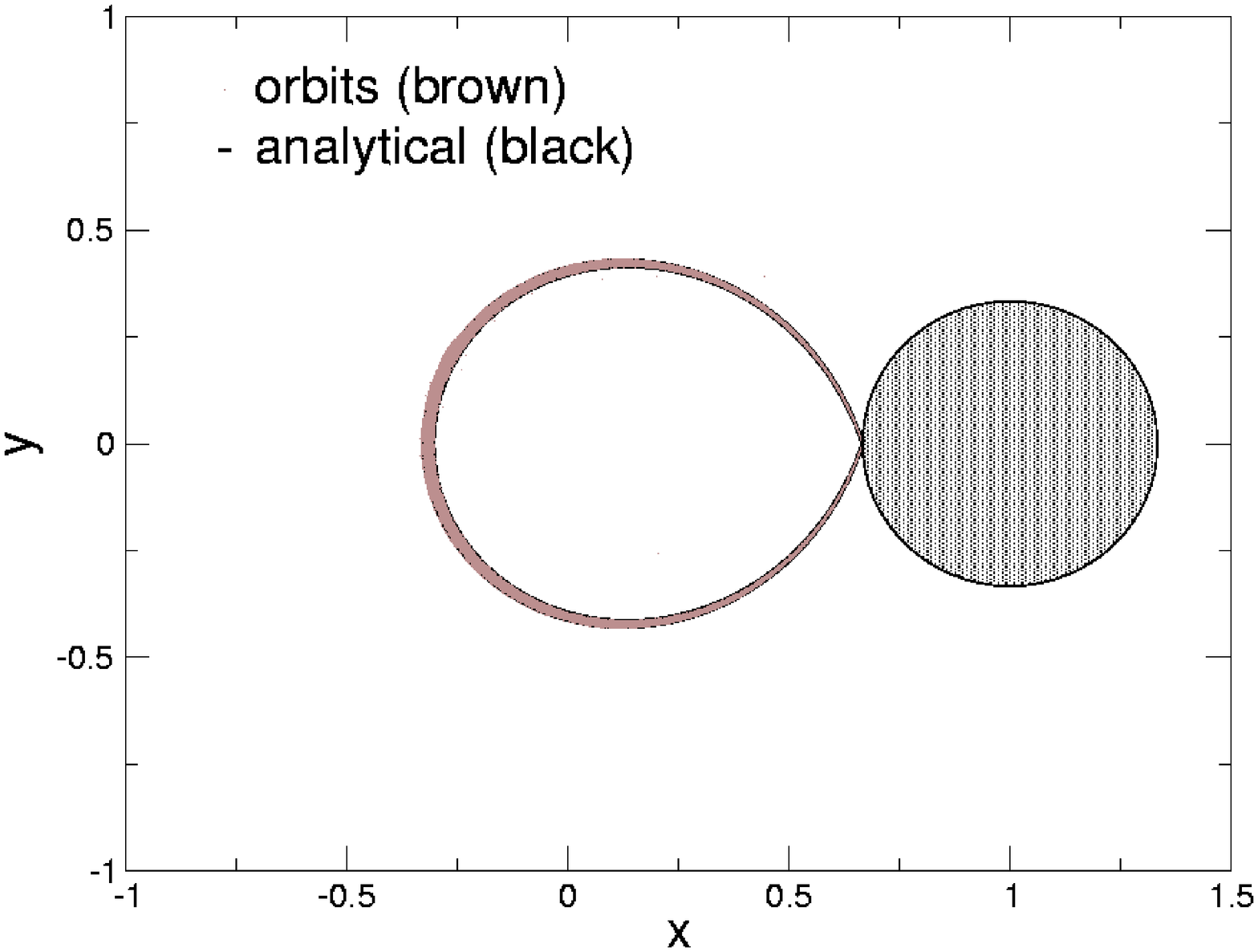} \hspace{0.5cm}
  \includegraphics[angle=270,width=7cm]{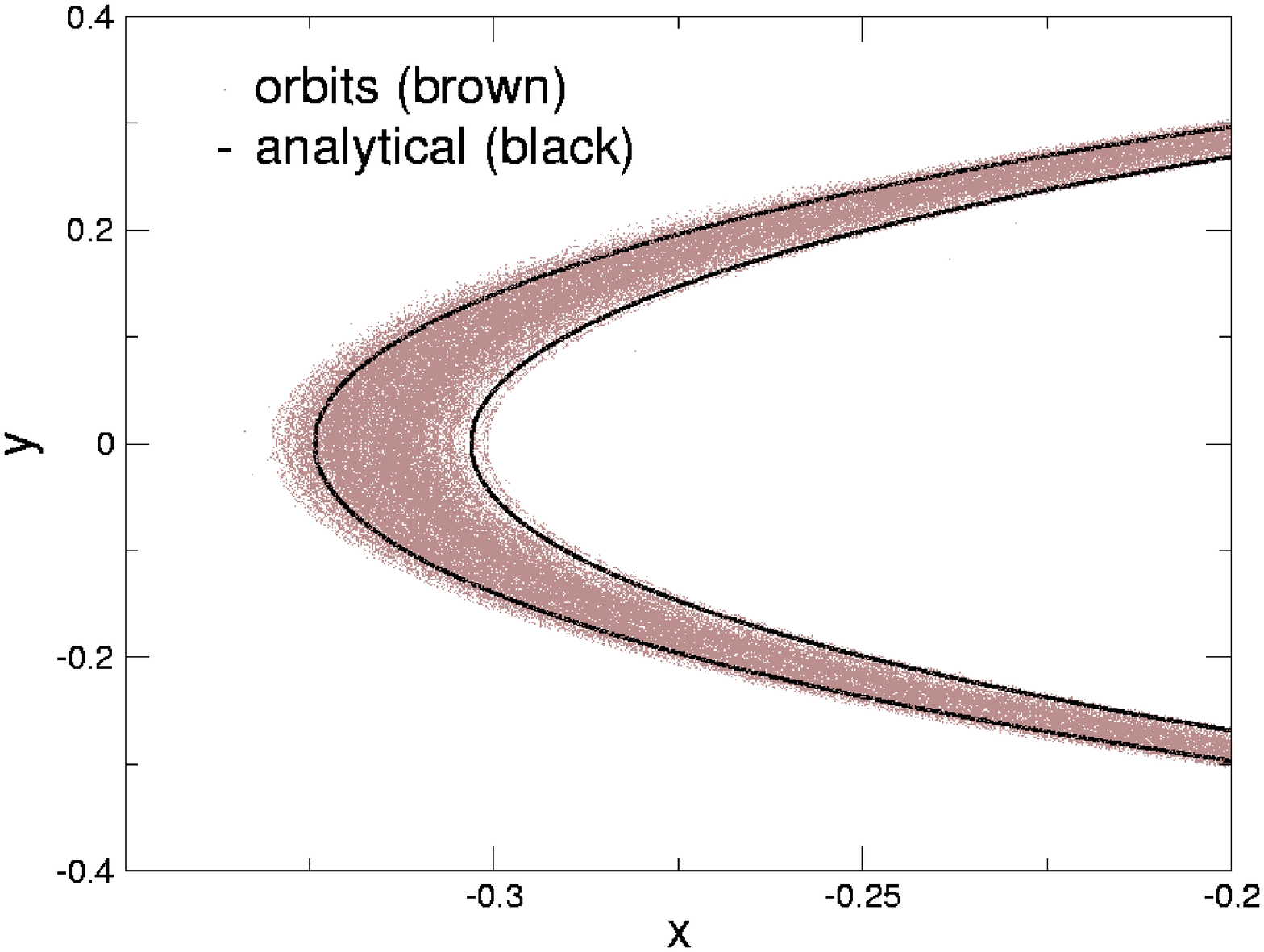}
\caption{\label{fig4} 
  (a)~Stable ring of non--interacting particles associated with the $n=0$
  stable region of the disk on a circular orbit. (b)~Zoom of a region of the
  ring. The red lines represent the limits defined by ${\rm Tr\,} D{\cal P}_J
  = \pm 2$. }
\end{figure*}

We focus on the latter set of initial conditions in phase space, which we
refer to as the particles of the ring, or ring particles. The motion of the
ring particles, as mentioned before, may be periodic, quasi--periodic or
chaotic. Therefore, we are not assuming or imposing any specific relation with
the period of the disk, $2\pi/\omega$, other than to assure consecutive
collisions. This is important because no resonance condition is implied: Even
in the case of periodic motion, the precise period of the radial collision
periodic orbits, $\Delta\phi/\omega$, may not be a rational of the period of
the disk. Of particular interest is certainly the pattern formed by this set
of particles in $X$--$Y$ space (inertial frame). Figures~\ref{fig4} show the
pattern obtained after few thousands of revolutions of the disk, which results
from a large number of particles whose initial conditions are close to the
maximum of the $J_{n=0}$ curve. The plot shows the $(X,Y)$ position of all
particles, in an inertial frame, that have not escaped at the time when the
``photograph'' was taken. Rigorously speaking, there may be still some
particles in this plot which do not belong to the region of trapped motion
although they have not escaped when the plot was made. We shall simply say
that the number of such particles is much smaller than the number of particles
that stay in the ring for ever. The plot justifies by itself to call this
pattern a {\it ring}.

In Fig.~\ref{fig4} we also plot the curves corresponding to ${\rm Tr\,} D{\cal
  P}_J = \pm 2$. These curves approximate the boundaries of the ring, which
follows directly from the stability arguments used for its construction. They
do not define true boundaries of the ring since they are related with the
appearance and destruction of the central elliptic fixed point; the satellite
structure around them defines some further thickness. Note that the nature of
these curves, i.e. the whole bifurcation scenario, implies naturally sharp
edges of the ring. It is also worth noticing the fact that the narrowness of
the ring is a consequence of the relatively small area occupied by the region
of trapped motion in phase space; this also takes into account the (small)
variations induced by changing $J$. Furthermore, the ring can be characterized
as eccentric, in the sense that there are two points, the periapse and
apoapse, forming a line that intersects the origin, the line of apsides. This
property is collective, in the sense that it is defined by the ensemble of
particles that builds the ring. We emphasize that these properties do not
depend on the particular interaction we have considered here: They hold for
any scattering system which displays stable motion confined to a tiny region
in phase space. This implies that the arguments hold under generic small
perturbations. Putting it differently, the phase--space structures in which we
are interested do display structural stability. The robustness of the argument
follows from the fact that the rings are constructed from phase--space
considerations which hold generically.

Further properties of the ring, which may not be generic as the preceeding
ones, are the following: For each situation in which the consecutive radial
collision orbits are stable, one such ring will be found. In the context of
the present example, this is so around each maximum and minimum of the curves
$J_n$. The ring particles move along the same direction of the disk for
negative $J$, while the motion is retrograde for $J>0$. The motion of the ring
is a rigid rotation around the origin, having the same period of the disk. All
rings share the same line of apsides. Finally, while it holds generically that
the rings are eccentric, the rather large eccentricity found in the present
example is surely an artifact of the toy model.

We shall finish this section commenting that our present approach only serves
to describe the occurrence of rings. We cannot distinguish whether the rings
appeared at the same time that the main planet and the shepherds, or whether
they formed after a catastrophic event like a collision with a big body. In
the first case, the ring particles are spread allover in phase space, and
those which remain within the ring are the ones within the regions of trapped
motion. We can consider the second option like a swarm of particles whose
initial conditions differ little and are defined on a small but closed region
in phase space, in particular in a small neighborhood in position space.
Again, the particles that remain in the ring are within the region of trapped
motion. However, they do not initially fill all this region: Time evolution
makes that they spread over the region of trapped motion in a short
time--scale and finally define a ring. This situation is similar to
thermalization processes in standard statistical mechanics.

\section{Rings in the elliptic case}
\label{sec4}

Here, we shall show that the same line of argumentation and the construction
described above can be carried on when the disk is moving an a Kepler ellipse.
We must emphasize that this is not a trivial generalization. We remark that
the Hamiltonian~\RefEq{eq2} can be interpreted as a two degrees of freedom
system with the addition of a quasi--periodic time--dependent perturbation.
Then, the system has effectively two--and--half degrees of freedom and the
Jacobi integral in the rotating--pulsating coordinates is no longer a
conserved quantity. Therefore, phase space is five dimensional and Poincar\'e
surfaces of section are of no practical use. Recent research on chaotic
scattering in many dimensions has revealed new interesting and intricated
behavior~\cite{Linos,Uzer02}. Furthermore, and more important, the argument
used to prove the existence of stable motion in the circular case relies on
the {\it generic} character of the saddle--center bifurcation. This is only
valid when a {\it single} parameter is varied; for the circular case this is a
consequence of the conservation of the Jacobi integral. However, for non--zero
eccentricity there is no constant of motion, and hence, the bifurcation
scenario involves at least two parameters.

\begin{figure}
  \includegraphics[angle=270,width=8.5cm]{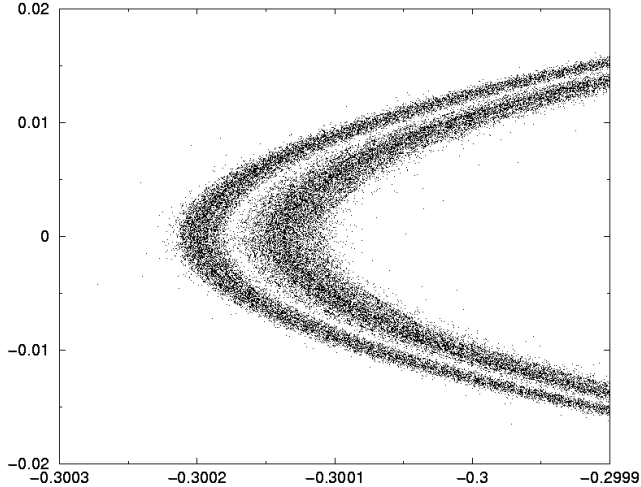}
\caption{\label{fig5} 
  Detail of the ring of non--interacting particles when the disk moves on an
  eccentric Kepler orbit ($\varepsilon=0.00165$). Notice the low density of
  particles found at the middle region of the ring.}
\end{figure}

As mentioned above, the elliptic motion introduces a kin of quasi--periodic
time perturbation on the circular case. It is therefore meaningless to look
for periodic motion, as we did in the circular case. We remark though
that the analytical continuation method~\cite{Siegel} can be used to prove the
existence of periodic orbits; these are deformations from a class of radial
collision orbits of the circular case~\cite{Benet04}. The important organizing
centers in phase space are in this case stable tori~\cite{Gab01}. While more
complicated situations may arise with respect to the stability of such
invariant structures in phase space, due to its larger dimensionality, we
shall mention that stable tori can indeed be found, at least for quite small
values of the eccentricity. Below we present evidence that supports this.
Close to these stable tori, the structure of phase space is somewhat similar
to the one discussed for the circular motion. Then, there is a region of
dynamical trapped motion due to consecutive collisions. The motion of
non--interacting particles started there can be quasi--periodic or chaotic.
Hence rings occur in the same way: Plotting at a given time the position of
the ensemble of non--interacting particles that are not ejected after a few
thousand collisions yields a ring, as shown in Fig.~\ref{fig5}. We notice that
the ring is also sharp--edged, eccentric and narrow: As mentioned above, these
properties hold generically. We remark that the ring in this case is narrower
than in the circular case. This may be interpreted as some destabilizing
effect due to the eccentricity of the shepherd moon. In Fig.~\ref{fig5} we
also observe a gap, or region of comparatively low density of particles, which
divides the ring in two components. We have no satisfactory explanation for
the occurrence of this region at the moment. While it is tempting to associate
this gap with different strands or ring components as those observed in
Satunrn's F ring, and its occurrence to the eccentricity of the shepherds,
more understanding on this aspect is certainly needed.

\begin{figure}
  \includegraphics[angle=270,width=9.5cm]{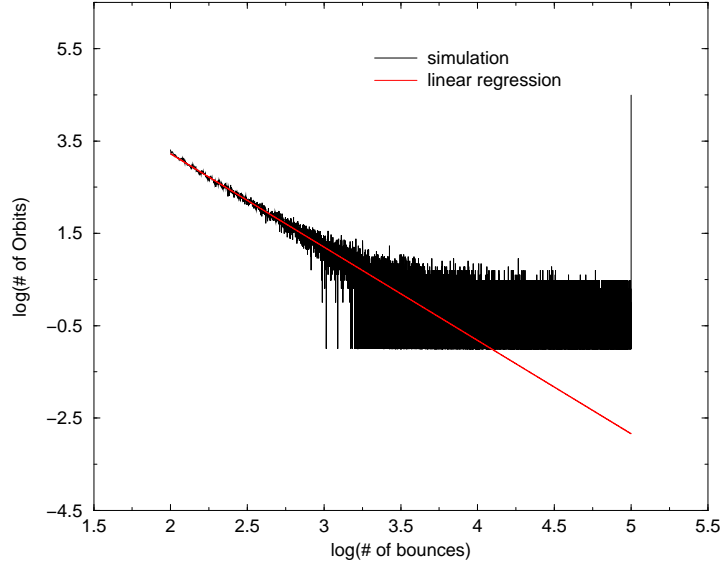}
\caption{\label{fig6} 
  Differential escape rate for $\varepsilon=0.00165$. The red line shows the
  results obtained by a linear regression. The peak at the right of the plot
  indicates the existence of particles that do not escape with an algebraic
  decay. These particles are dynamically trapped. }
\end{figure}
We present now some evidence that pure stable quasi--periodic tori indeed
exist. First, we mention that we have found trajectories that survive more
than 100000 collisions. While this number of bounces is quite large in the
numerical calculations, these trajectories may still not be trapped due to the
large dimensionality of phase space, in particular if they display chaotic
motion. Indeed, Arnold diffusion~\cite{Arnold64,Lega03} could be relevant in
the present situation. Yet, the typical time scale in which Arnold diffusion
takes place is exponentially large and thus our simulations have not detected
it. Secondly, we investigate the escape rate of an ensemble of initial
conditions near a region where we suspect there is trapped motion. The escape
rate of a hyperbolic system shows an exponential decay in contrast to the
algebraic one observed for a non--hyperbolic system~\cite{Ding90}. For the
latter, the exponent is close to 2. We thus expect to have an algebraic decay,
with non--statistical deviations on the tail if there are initial conditions
within a region of trapped motion. In Fig.~\ref{fig6} the differential escape
rate is shown for $\varepsilon=0.00165$, showing the expected (initial)
algebraic decay with exponent $\approx 2.2$. The large peak that appears in
the tail at the cut--off in the number of bounces indicates that there is
non--zero overlap between the region where the initial conditions were chosen,
and the region of stable trapped motion associated with a central stable tori.
Therefore, a stable trapped motion does exist for this eccentricity, which
yields the ring shown in Fig.~\ref{fig5}. Notice that this method, while it
does not provide any information of the size in phase space of the trapped
region or on dynamics, shows that the measure of the set is strictly
non--zero.

\section{conclusions and outlook}
\label{concl}

In this paper we have discussed generic mechanisms in phase space for the
occurrence of narrow eccentric planetary--like rings due to the interaction
with shepherd moons. Our approach is based on a Hamiltonian formulation and
assumes non--interacting particles. The non--interacting character of our
approach is only a first approximation in the same sense that the mean--field
theory is for nuclear physics. We have used as a simple but unrealistic
example for the argumentation: An open (scattering) billiard system on a
Kepler orbit. We have focused on the structure of phase space, and extended
previous results to the case of an eccentric Kepler orbit. Although the
general arguments carry over, the generalization is not straightforward: The
enlarged dimensionality introduces new possibilities in the stability
properties of the fixed points and on the bifurcation scenario involving now
at least two parameters. We find that the eccentric character of the orbit
indeed influences the structure of the ring, which turns out to be narrower in
comparison to the circular counterpart and also displays a gap. We have no
satisfactory explanation for the latter aspect yet, and do not know whether it
is an artifact of our model example or something with deeper significance. We
have also provided a more detailed analytical framework to study the circular
case, in particular providing estimates for the width of the ring. We
emphasize that these new results are by no means a trivial extension: The
dimensionality of the system implied through the time--dependence makes this a
subtle task. To find rings in the present toy example which has no attractive
interaction shows the genericity of the arguments.

The relevant structures in phase space correspond to those of pure trapped
motion, which are organized around stable periodic orbits, or more generally,
around stable tori. In phase space these regions are bounded by the invariant
manifolds of unstable orbits. Within these regions no escape is possible. If
there exists such a set of pure bounded motion in phase space, it is strictly
of non--zero measure.  An ensemble of (non--interacting) particles defined by
their initial conditions within these regions yields in position space a ring.
The motion of particles with initial conditions in these regions must be
contrasted with those outside: In the latter case, the particles are ejected
from the system after few periods of rotation. This situation leads naturally
to sharp--edges of the ring. This scenario is {\it robust}, in the sense that
it does not depend on the particular interaction between the particles of the
ring, the shepherd moons and the central planet: The rotation of the shepherd
moon by itself can build the conditions to have trapped motion. Once there is
pure trapped motion the existence of such a ring follows immediately. That is,
within a more realistic description, non--keplerian effects like planet
oblateness can also be accounted for. The rings obtained display sharp--edges,
are eccentric and narrow. We emphasize that these properties do not assume any
specific resonance condition between the particle motion and the shepherd
moons, as it is often considered or assumed. It is the existence of stable
motion and its organizing centers in phase space what determines the existence
of such a ring. We believe that this scenario may provide an explanation to
the observed feature of apse alignment, which remains an open problem within
the usual shepherding model.

This example can be taken over to consider the interaction with two disks on
non--overlapping keplerian orbits, in order to model the more realistic
situation of two shepherd moons. Using the same ideas, it is possible to
define a (geometrical) criterion essentially given by the intersection of the
ring (due to the outer disk) and the orbit of the inner one. This criterion
may display a specific dependence on the actual position of the outer ring in
the case of eccentric orbits. A possible generalization to the case with
smooth potentials and in particular the $1/r$ case~\cite{Benet01}, should
maybe incorporate the results related to the stability of the organizing
center, and then provide estimates of the width of the ring.

Interparticle interactions do indeed play a crucial role in the dynamics of
narrow rings. Our current point of view is that they can be incorporated
within the present modeling, at least for a dilute ring, as some chaotic
diffusion (Brownian motion) in phase space, with probably some constrains.
Such diffusive process induces small fluctuations in the value of the Jacobi
integral and in the instantaneous direction of motion of the particles. One
may thus impose that such a collision process conserves the total angular
momentum of the colliding particles. This will effectively make that the
dimension of phase space incorporates the number of particles. Future
research will be carried along this direction.

\begin{acknowledgements}
  We would like to thank C. Jung and T.H. Seligman for helpful discussions. We
  acknowledge financial support provided by the projects IN--101603
  (DGAPA--UNAM) and 43375 (CONACyT) as well as from the Swiss National
  Foundation.
\end{acknowledgements}

\end{document}